# Enabling Clinical Use of Linear Energy Transfer in Proton Therapy for Head and Neck Cancer— A Review of Implications for Treatment Planning and Adverse Events Study

**Authors:**

Jingyuan Chen, PhD[1], Yunze Yang, PhD[2], Hongying Feng, PhD[1,3,4], Chenbin Liu, PhD[5], Lian Zhang, PhD[1,6], Jason Holmes, PhD[1], Zhengliang Liu, MS[7], Haibo Lin, PhD[8], Tianming Liu, PhD[7], Charles B. Simone II, MD[8], Nancy Y. Lee, MD[9], Steven E. Frank, MD[10], Daniel J. Ma, MD[11], Samir H. Patel, MD[1], Wei Liu, PhD[1]

[1]Department of Radiation Oncology, Mayo Clinic, Phoenix, AZ 85054, USA

[2]Department of Radiation Oncology, the University of Miami, FL 33136, USA

[3]College of Mechanical and Power Engineering, China Three Gorges University, Yichang, Hubei 443002, People's Republic of China

[4]Department of Radiation Oncology, Guangzhou Concord Cancer Center, Guangzhou, Guangdong, 510555, People's Republic of China

[5]Cancer Hospital & Shenzhen Hospital, Chinese Academy of Medical Sciences and Peking Union Medical College, Shenzhen, China

[6]Department of Oncology, The First Hospital of Hebei Medical University, Shijiazhuang, Hebei, 050023, People's Republic of China

[7]School of Computing, the University of Georgia, Athens, GA 30602, USA

[8]New York Proton Center, New York, NY 10035, USA

[9]Department of Radiation Oncology, Memorial Sloan Kettering Cancer Center, New York, NY 10065, USA

[10]Department of Radiation Oncology, The University of Texas MD Anderson Cancer Center, Houston, TX 77030, USA

[11]Department of Radiation Oncology, Mayo Clinic, Rochester, MN 55905, USA

Corresponding author: Wei Liu, PhD, Professor of Radiation Oncology, Department of Radiation Oncology, Mayo Clinic Arizona; E-mail: Liu.Wei@mayo.edu.

## Conflicts of Interest Disclosure Statement

No

## Funding Statement

This research was supported by the National Cancer Institute (NCI) R01CA280134, the Eric & Wendy Schmidt Fund for AI Research & Innovation, The Fred C. and Katherine B. Anderson Foundation, and the Kemper Marley Foundation.

## Acknowledgments

This research was supported by the National Cancer Institute (NCI) R01CA280134, the Eric & Wendy Schmidt Fund for AI Research & Innovation, The Fred C. and Katherine B. Anderson Foundation, and the Kemper Marley Foundation.

# Abstract

Proton therapy offers significant advantages due to its unique physical and biological properties, particularly the Bragg peak, enabling precise dose delivery to tumors while sparing healthy tissues. However, the clinical implementation is challenged by the oversimplification of the relative biological effectiveness (RBE) as a fixed value of 1.1, which does not account for the complex interplay between dose, linear energy transfer (LET), and biological endpoints. Lack of heterogeneity control or the understanding of the complex interplay may result in unexpected adverse events and suboptimal patient outcomes. On the other hand, expanding our knowledge of variable tumor RBE and LET optimization may provide a better management strategy for radioresistant tumors.

This review examines recent advancements in LET calculation methods, including analytical models and Monte Carlo simulations. The integration of LET into plan evaluation is assessed to enhance plan quality control. LET-guided robust optimization demonstrates promise in minimizing high-LET exposure to organs at risk, thereby reducing the risk of adverse events.

Dosimetric seed spot analysis is discussed to show its importance in revealing the true LET-related effect upon the adverse event initialization by finding the lesion origins and eliminating the confounding factors from the biological processes. Dose-LET volume histograms (DLVH) are discussed as effective tools for correlating physical dose and LET with clinical outcomes, enabling the derivation of clinically relevant dose-LET volume constraints without reliance on uncertain RBE models. Based on DLVH, the dose-LET volume constraints (DLVC)-guided robust optimization is introduced to upgrade conventional dose-volume constraints-based robust optimization, which optimizes the joint distribution of dose and LET simultaneously.

In conclusion, translating the advances in LET-related research into clinical practice necessitates a better understanding of the LET-related biological mechanisms and the development of clinically relevant LET-related volume constraints directly derived from the clinical outcomes. Future research is needed to refine these models and conduct prospective trials to assess the clinical benefits of LET-guided optimization on patient outcomes.

# Introduction

Radiation therapy (RT) is a standard treatment option used for 50-75% of cancer patients (1-3). Over recent decades, proton therapy has seen significant technological advancements and increased clinical applications(4,5). The proton beam is characterized by its Bragg peak, which has a sharp dose fall-off after the target. This characteristic allows proton therapy to achieve a lower entrance to peak dose ratio, improved dose conformality to the target and enhanced dose protection to organs at risk (OARs) compared to conventional photon therapy(6-13).

Despite the dosimetric benefits, proton therapy faces a major challenge of relative biological effectiveness (RBE) (14-19). In contrast to photons, protons impart most of their energy over a short distance, and thus, induce high linear energy transfer (LET) near the distal end of the Bragg Peak. Hence, the biological effect of proton therapy should be determined by both dose and LET (and possibly other factors) (14,15,20-23). Various studies on *in vitro* cell experiments(24,25) show that RBE increases with elevated LET, while clinical outcome data are less clear regarding the impact of LET on RBE(26-39). An RBE >1.1 for adverse events (AEs) associated with higher LET within OARs has been reported for rib fracture(40), rectal bleeding(41), mandible osteoradionecrosis(42,43), brain necrosis(26,28,34,44), and late-phase pulmonary changes(29) in cancer patients treated with proton therapy. An improved understanding of the relationship between physical dose, LET, and AEs in proton therapy planning is greatly needed to improve treatment planning.

Several phenomenological and mechanistic RBE models have been developed to calculate RBE from LET and physical dose (45-54). However, systematic evaluations have shown that in vitro RBE predictions can vary greatly across different models (55). This significant variability is largely due to the use of tissue-specific α/β ratios in these models, which themselves can have significant parameter uncertainties (46,56). Moreover, substantial discrepancies have been reported between in vitro and in vivo RBE results (24). Since outcomes from clonogenic assays do not necessarily reflect the clinical responses of cancer patients undergoing proton therapy, current RBE models are hindered by considerable biological and parametric uncertainties, limiting the clinical application of LET.

In clinical practice, a fixed RBE value of 1.1 represents higher cell-killing effect compared to photons. Proton therapy planning typically relies solely on dose calculations and overlooks critical LET information as well as variable RBE of tumors based on histology and fraction size(17,18). This oversimplification has adversely affected the efficacy of proton therapy, leading to unexpected AEs that place additional burdens on the healthcare system and increase financial costs (20,57-59). Therefore, there is an urgent need to incorporate LET considerations into plan optimization and evaluation to reduce AEs.

In this paper, we first review and summarize the current research on LET calculation, LET-guided plan evaluation, and LET-guided plan optimization. Then, we discussed the most recent developments in LET-related AE studies, with particular focuses on dosimetric seed spot analysis, dose-LET volume histogram (DLVH), and how use dose LET volume constraint-based robust optimization is used prospectively to adjust the dose and LET distribution simultaneously potentially to minimize the incidence rates of AEs.

## LET calculation

Analytical calculations and Monte Carlo simulations are two main methods to calculate the LET. The analytical LET calculation methods(47,60-67) have been used in clinical practice owing to their high efficiency, acceptable computational accuracy, and other historical reasons. One-dimensional LET models(45,60,61) assumed uniform lateral LET(64-66). Three-dimensional LET calculation models considered lateral LET variations(64-66). Monte Carlo (MC) simulations (68-73) typically offer greater accuracy than analytical methods, especially in inhomogeneous geometries, but they require significantly longer computation times, particularly when general-purpose MC algorithms(69,74-76) are used. Fast MC codes(77-83) have been developed and clinically implemented, speeding up proton dose calculations using simplified physics models, GPU acceleration or combined. Moreover, commercial treatment planning systems like RayStation (RaySearch Laboratories, Stockholm, Sweden) (84) and Eclipse (Varian Medical Systems, Palo Alto, CA, USA) (85,86) now feature fast MC capabilities for routine dose calculations(87-90). However, despite the progress in MC-based robust optimization (91,92) and robustness evaluation(93-95), LET calculations based on fast MC have yet to be incorporated into any commercial TPSs for clinical use. Recently, the deep learning-based dose and LET calculation engines were also proposed (96,97).

## LET-guided plan evaluation

Studies have found a strong correlation between high dose and high LET distribution in OARs and AEs(42,43). The lack of accountability of variable and high LET distributions in clinical practice may result in severe AEs and undesirable patient outcomes in proton therapy. LET-guided dosimetric evaluation has become more common in proton therapy centers, serving as a biological effect evaluation tool for intensity-modulated proton therapy (IMPT) plans (98,99). A recent survey showed that 16 of 25 European proton centers called for more retrospective or prospective outcome studies, investigating the effect of variable RBEs induced by high LET, and 18 centers called for LET and RBE calculation and visualization tools(100). Typically, LET distributions of IMPT treatment plans are generated through analytical calculations or Monte Carlo simulations for further review. To assess the biological effects of an IMPT plan, physicians or physicists may examine areas where high doses and high LET overlap, aiming to minimize such overlaps in critical structures, or they may analyze the biological dose distribution based on dose and LET using various RBE models. In some centers, like the Mayo Clinic in Arizona, LET-guided plan evaluation is now a routine process for all patients undergoing IMPT, while at other centers, LET evaluation and optimization are performed on a more ad-hoc basis.

## LET-guided plan optimization

MC simulations and experiments on water phantom have shown that similar dose distributions can lead to significantly different LET distributions(101). Therefore, during treatment planning, it is necessary to optimize LET to reduce potential AE risks (102,103). Building on top of dose optimization algorithms, various LET/RBE-guided plan optimization approaches have been developed. Some of these algorithms directly use LET in the objective function(25,104-107), whereas others use LET-related terms indirectly in the optimization process(15,106,108,109). However, in LET-guided optimization, it is crucial to balance the trade-off between the LET and dose distributions during the optimization process to ensure that optimizing LET does not compromise overall plan quality (110).

Robust optimization (RO) is common in proton therapy(87,111-116). Proton therapy is highly sensitive to range and setup uncertainties, especially IMPT (117-121), and RO can generate robust plans by accounting for either the voxel-wise or objective-wise worst-case scenarios during optimization (11,12,106,112,122-140). LET-guided robust optimization has also been developed, which generally incorporates LET/RBE-related constraints for OARs in the dose-based robust optimization and adds additional LET/RBE-related penalty terms in the objective function (105,106,109,141-146). These methods have improved the LET distribution in OARs while maintaining comparable plan quality and robustness.

The LET peak of proton beams occurs beyond the Bragg peak; therefore, optimizing proton beam angles and spot locations to deposit the LET peak at less hazardous regions while keeping the dose peaks not moved can result in a superior LET distribution (147-153). However, most current optimization methods focus solely on beam weight optimization. Incorporating beam angle optimization may be more beneficial for treatment plans in anatomically complex regions, but beam angle optimization will significantly increase the computation time (104,109,146). Additionally, increasing the number of beams can also optimize LET distribution in some patients (31,154). Spot-scanning proton arc therapy (SPArcT) with infinite beams can achieve superior dose and LET target conformity(155). However, further research is needed to establish its clinical advantages (156-159).

## Studying LET-related clinical outcomes

Two main approaches are currently used in studying LET-related clinical outcomes. One approach is voxel-based analysis, which compares regions exhibiting AEs with a matched healthy region. (26,28,30,31,160-162). The second approach is to study AEs at the organ level, which involves a population-based analysis that establishes a relationship between the epidemiological probability of complications and the treatment modality, whether the difference involves photon or proton therapy. (29,37,43)

In the first approach, individual voxels are utilized as data points for analysis. The potential correlation between clinical outcomes (such as whether a voxel is damaged) and both dose and LET was investigated. This approach was based on two fundamental assumptions: (1) all damaged voxels are a result of dosimetric effects, namely dose and LET; (2) within the AE regions of the same patients, voxels are considered to be independent of one another. However, these assumptions are not universally applicable. Clinical observations suggest that the AE region will expand/shrink over time because of biological processes. Voxels in one single lesion are not fully independent from each other. In addition, the potential volume effect is not considered in this approach. The established normal tissue complication probability (NTCP) is at voxel-level and only considers the dose and/or LET values.

In the second approach, organ-level NTCP is established by comparing clinical outcome differences between photon and proton patient cohorts. Although there are indications of increased RBE with protons, the quantification is solely based on dose-volume metrics. The precise contribution from LET is challenging to assess, not only due to the absence of meaningful LET quantification at the organ level, but also because LET is highly heterogeneous. Its distribution within the organ and its relation to dose distribution matter.

## Novel approaches for LET-based outcome studies

### Dose-LET volume histogram in the organ-level adverse event analysis

Dose-LET volume histogram (DLVH) is a novel tool for studying the combined effects of the dose and LET on patient outcomes. As shown in Fig.1, DLVH is a cumulative volume histogram tool following a similar statistical concept as a dose volume histogram (DVH) (41). Different from a 2D plot of DVH, DLVH is a 3D surface plot. Well-defined physical quantities - dose (Gy) and LET (keV/μm) - are constructed as two horizontal axes, whereas the third vertical dimension shows the normalized volume. With DLVH, well-defined physics quantities can be associated, such as dose, LET, and volume of OARs with AEs. Clinically relevant dose-LET volume constraints (DLVCs) can be obtained(41) without the inclusion of RBE models(41) to bypass the uncertainties in the current RBE models.

The DLVH approach offers several advantages: 1) It accounts for the volume effect, beyond just the numerical values of dose and LET; 2) It is possible to establish the relation of outcome versus the DLVH index, $V(d, l)$, the specific volume having both a certain dose and a certain LET. Thus, the interplay of dose and LET distributions within the organ will be considered; 3) Instead of incorporating assumed dose-LET relations as variables into the regression analysis, it is possible to derive the dose-LET relations based on patterns observed from the regression analysis; and 4) DLVH maintains the integrity of LET information. DLVH analysis allows one to use patient cohort data at the organ level while precisely investigating the LET contribution. This avoids the data independence issue in the voxel-based analysis.

Based on DLVH, DLVC-guided robust optimization has been proposed as an efficient method to simultaneously control 3D dose and LET distributions during proton therapy treatment planning(163). This method upgrades proton therapy treatment planning from 2D DVH-based (164,165) to 3D DLVH-based by considering dose, LET, and volume, and implements DLVCs as soft constraints in the objective function(166-168), thereby effectively reducing the number of potential seed spots and lowering the incidence of corresponding AEs.

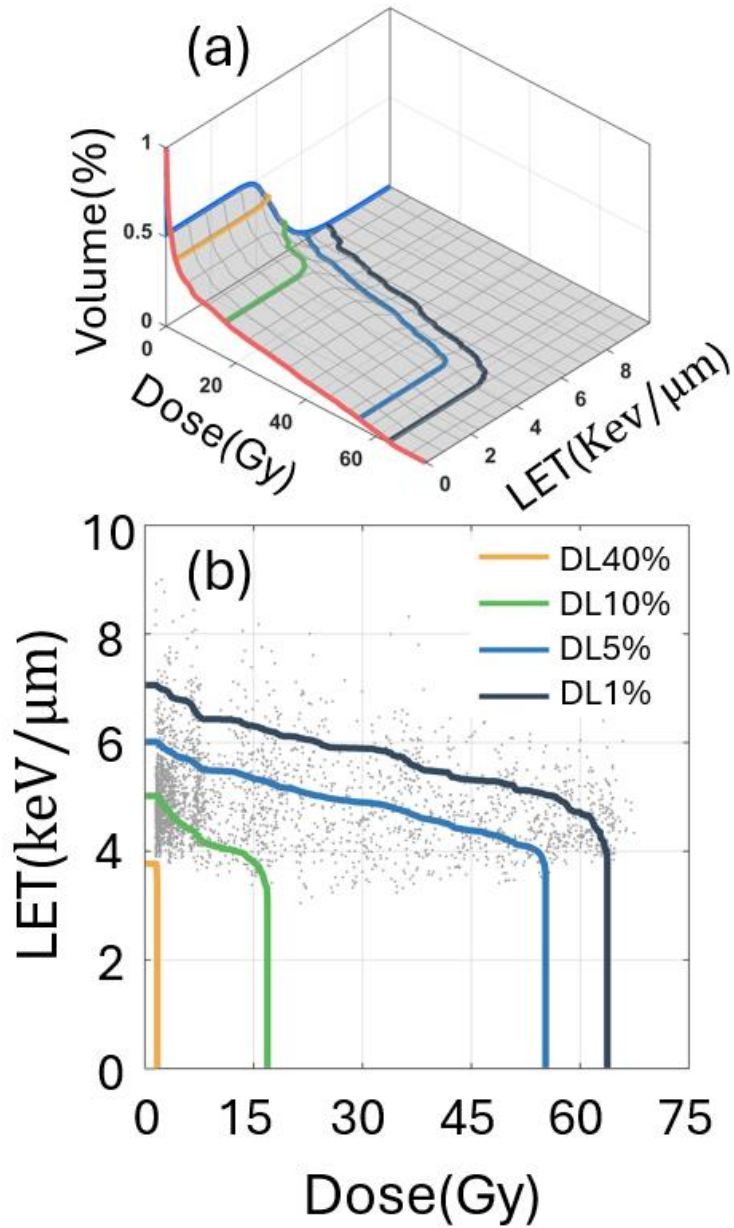

Figure 1: Sketches about the dose linear-energy-transfer (LET) volume histogram (DLVH). (a) Three-dimensional DLVH surface. The solid lines on the surface are the iso-volume contour lines DL$v$%. (b) The projected two-dimensional DLVH of (a) and the iso-volume contour lines. The gray dots represent the voxels of the structure.

**Dosimetric seed spot analysis in the voxel-level adverse event analysis**

As mentioned above, the AE sites progress over time. Once radiation damage is initialized by dose and LET (i.e., radiation effect), biological processes may take over and the original AE sites may expand spatially(43). Analysis using all voxels, especially those low dose/LET voxels in the expanded AE sites, may have masked the importance of the radiation effect that triggered the AEs initialization (26,28,32,39,43,44,169). In addition, in such analysis(28,32,34,169,170), each voxel was treated as an independent data point to establish the relationship of dose and LET with the patient outcome. This approach breaks the fundamental assumption that "any regression methods require independent data points" (26).

Recently, dosimetric seed spot analysis proposed a spatial clustering method to eliminate "noises" from biological processes to study AE initialization. As illustrated in Fig. 2, this approach finds several clusters (seed spots), each representing a spatially independent lesion origin. Although it is impossible to fully get rid of the biological impacts, reducing the number of data points for analysis and finding their independent representatives improves the data independence and reduces the noise from the overrepresentation of non-contributing voxels. Based on dosimetric seed spot analysis, some research(42,43) suggested that RBEs are underestimated in current clinical practice and the LET-enhancing effect is critical for AE initialization in head and neck patients.

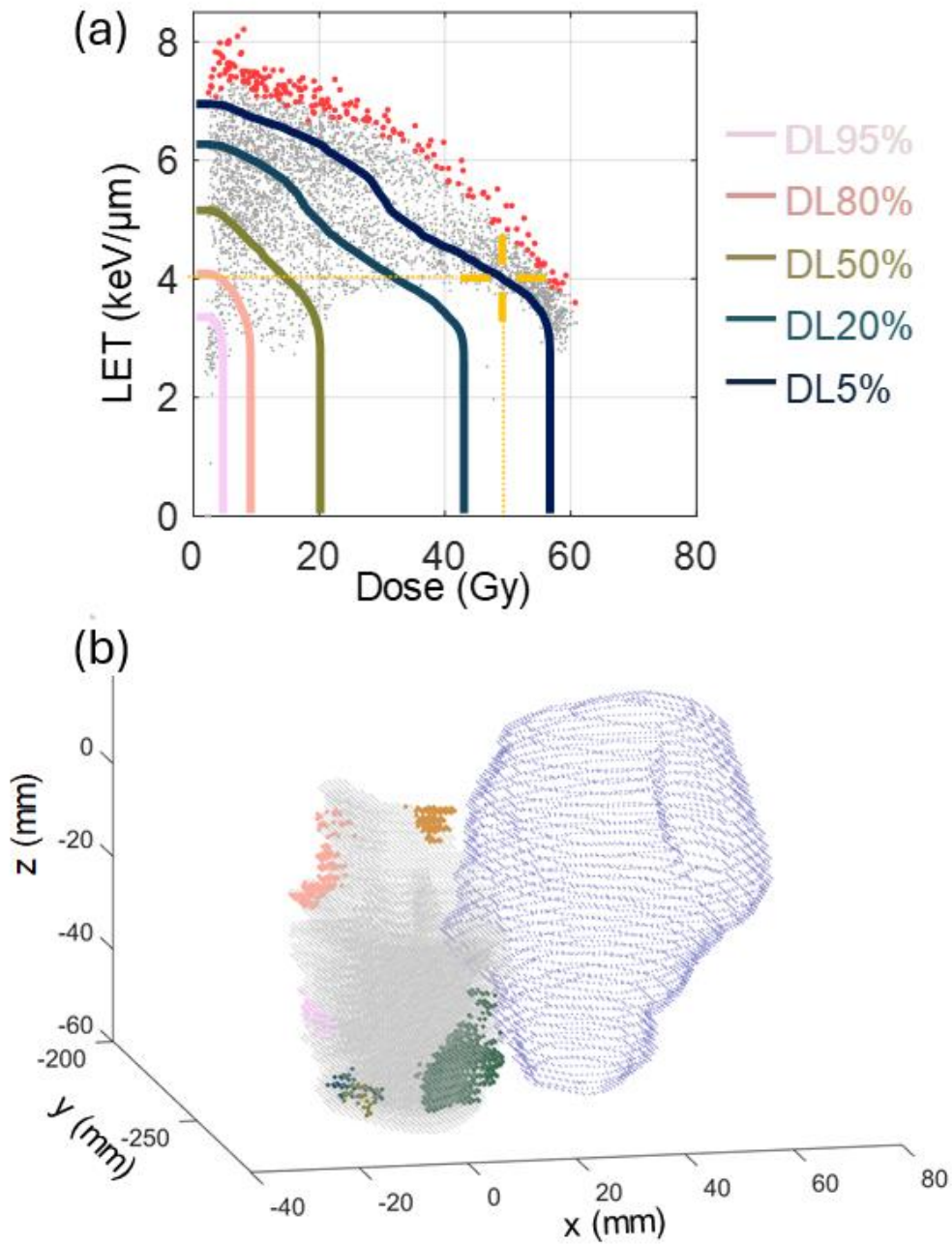

Figure 2: An example of the dosimetric seed spot analysis in the voxel-level adverse event analysis. (a) Dose linear-energy-transfer (LET) volume histogram (DLVH) of an adverse event (AE) region in one head-and-neck (H&N) patient. The grey dots represent voxels of the structure. The solid lines are the iso-volume contour lines DL$v$% of the DLVH. The assumed critical voxels for seed

spot analysis are represented by the red dots, which are the highest 5% LET voxels selected from each dose bin within the moderate to high dose range. Potential voxels influenced by biological effects in in-field AE regions with low doses and low LET are enclosed within the light blue dashed oval. The purple dashed circle and green dashed oval respectively denote possible voxels in in-field AE regions typically characterized by high doses, and out-of-field AE regions typically characterized by high LET. (b) Identification of seed spots within an AE region. The spatial distribution of seed spots is shown for mandibular osteoradionecrosis in a representative patient. Critical voxels in DLVH are identified and grouped into two seed spots, each highlighted in a different color. Other AE voxels are depicted in gray, while the high-dose clinical target volume (CTVHigh) is shown in blue. The figure is presented in DICOM coordinates.

## Conclusion

With the development of hardware and software in proton therapy, LET-related research has recently made significant progress, including studies on precise LET calculation, LET-guided plan evaluation, LET-guided plan optimization, and LET-related patient outcomes. However, LET-guided plan evaluation and optimization require a better understanding of the LET-related biological mechanisms and clinically relevant LET-related volume constraints directly derived from the clinical outcomes, both of which need further research.